\documentstyle[amsfonts,aps,prl]{revtex}
\draft
\begin{document}
\title
 { Canonical Nonabelian Dual Transformations in Supersymmetric Field Theories}
\author{ Thomas Curtright\cite{authTLC} }
\address{Department of Physics, University of Miami, Box 248046,
          Coral Gables, Florida 33124, USA}
\author{ Cosmas Zachos\cite{authCKZ} }
\address{High Energy Physics Division, Argonne National Laboratory,
          Argonne, Illinois 60439-4815, USA}
\date{February 21, 1995}
\maketitle

\begin{abstract}
A generating functional $F$ is found for a canonical nonabelian dual
transformation which maps the supersymmetric chiral O(4) $\sigma$-model
to an equivalent supersymmetric extension of the dual $\sigma$-model.
This $F$ produces a mapping between the classical phase spaces
of the two theories in which
the bosonic (coordinate) fields transform nonlocally,
the fermions undergo a local tangent space chiral rotation,
and all currents (fermionic and bosonic) mix locally.
Purely bosonic curvature-free currents of the chiral model
become a {\em symphysis} of purely bosonic and fermion bilinear currents
of the dual theory.
The corresponding transformation functional $T$
which relates wavefunctions in the two quantum theories
is argued to be {\em exactly} given by $T=\exp(iF)$.
\end{abstract}
\pacs{ hep-th/9502126 \qquad Miami TH/1/95 \qquad ANL-HEP-PR-95-14}
\widetext

In a general sense, a dual transformation is a map between two models
which preserves dynamical features and thereby facilitates our
understanding by providing alternate ``dual'' views of these features.
Examples include Kramers-Wannier duality, relating high and low
temperature limits of  lattice models;
bosonization of fermions, to reveal properties of solitons\cite{Mandelstam};
alternate Lorentz tensor descriptions of spinning fields; mirror symmetries
relating compactified spacetimes in string theories\cite{alvarez-gaume}; and
recent non-perturbative methods in supersymmetric four dimensional field
theories\cite{wittenseiberg}.

Here, we generate nonabelian field theory examples of dual transformations for
nonlinear $\sigma$-models in two spacetime dimensions, within the framework of
equal-time canonical transformations, in parallel to earlier abelian work for
the quantum Liouville theory\cite{CurtrightGhandour}. We have already
illustrated this equal-time mapping approach for purely bosonic
$\sigma$-models\cite{CurtrightZachos'94}.  We first summarize and clarify those
earlier results by comparing to the lagrangean framework. We then construct new
mappings by including fermions supersymmetrically. Our discussion concentrates
on classical field theory, but we believe an acceptable route to quantization
exists for these $\sigma$-models exactly as it does for the Liouville case: by
simply exponentiating the classical generating functional, which we discuss in
concluding.

For the supersymmetric case, our new dual transformation does not separately
map bosons to bosons and fermions to fermions; rather it ties together the
effects of both fermions and bosons in one model to produce the effects of
either bosons or fermions separately in the other, dual model.  This occurs
even though the map does not interchange fermions with bosons (e.g.~as in the
soliton studies mentioned). While this phenomenon might be prefigured in the
current identifications of canonical bosonization schemes\cite{Mandelstam}, we
believe its novelty warrants introduction of the term ``boson-fermion
symphysis''.

To begin, recall the standard bosonic
Chiral Model (CM) on O(4)~$\simeq$~O(3)$\times$O(3)$\simeq$SU(2)$\times$SU(2).
Expressed in geometrical terms, we have
${\cal L}_{CM}=\frac 12 g_{ab} \partial _\mu
\varphi ^a\partial ^\mu \varphi ^b$, where $g_{ab}$ is the metric on
the field manifold (three-sphere).
Explicitly, with group elements parameterized as
$U=$  $\varphi ^0+i\tau ^j\varphi ^j$,      $(j=1,2,3)$, where
$(\varphi ^0)^2+{\varphi }^2=1$,
and ${\varphi }^2\equiv $ $\sum_j(\varphi ^j)^2$, we may resolve
$\varphi ^0=\pm \sqrt{1-{\varphi }^2}$, to obtain
$g_{ab}= \delta ^{ab}+\varphi^a\varphi^b/(1-\varphi^2) $.
We re-express the action in terms of tangent
quantities through the use of either left- or right-invariant vielbeine
(e.g.~\cite{geometrostasis}).
Either choice yields the Sugawara current-current form
${\cal L}_{CM}=\frac 12 J_\mu^j J^{\mu ~j}$.
Choosing left-invariant dreibeine gives ``$V+A$'' currents which are
vectors on the tangent space.  In terms of the above explicit coordinates,
$U^{-1}\partial_{\mu} U= -i~ \tau^j J_\mu^j$,
$~J^i_\mu = -v_a^{~i} \partial_\mu \varphi^a$, where
$v_a^{~j}=\sqrt{1-\varphi^2} ~g_{aj} + \varepsilon^{ajb} \varphi^b~$.
Note that these currents are pure gauge, or curvature-free, such that
$ \varepsilon ^{\mu \nu }\left( \partial _\mu J^i_\nu +\varepsilon
^{ijk}J^j_\mu J^k_\nu \right) =0$.

In \cite{CurtrightZachos'94} we canonically mapped this model onto a
dual $\sigma$ model (DSM, with variables $\Phi^j$)
using the tangent space\cite{orlando} generating functional
$F[\Phi,\varphi]=\int~dx~\Phi^j J^{1~j}[\varphi]$.
Although we originally constructed $F$
in the hamiltonian framework by some indirect reasoning,
its structure is directly
evident within the lagrangean framework as follows.
Treating $J$ as independent
variables in ${\cal L}_{CM}=\frac 12 J_\mu^j J^{\mu ~j}$, we
impose the pure gauge condition by adding a Lagrange multiplier term,
${\cal L}_{\lambda}=\Phi^j~\varepsilon^{\mu \nu}
(\partial_\mu J_\nu ^j +\varepsilon^{jkl} J_\mu ^k J_\nu ^l)$. Then,
we complete a square for the $J$'s and eliminate them
from the dynamics in favor of the DSM\cite{Fridling}.
But to do this, we must first write
${\cal L}_{\lambda}=\partial_\mu (\Phi^j~\varepsilon^{\mu \nu}~J_\nu ^j)
- \varepsilon^{\mu \nu}~J_\nu ^j ~\partial_\mu \Phi^j~
+\varepsilon^{jkl}~\varepsilon^{\mu \nu}~ \Phi^j J_\mu ^k J_\nu ^l$.
The total divergence term, integrated over a world-sheet with
a constant time boundary, gives precisely our generating functional
relating the CM to the DSM.

The supersymmetric extension (SCM) of the CM is\cite{witten}:
${\cal L}_{SCM}=
\frac 12 g_{ab}~\partial _\mu \varphi ^a
\partial ^\mu \varphi ^b+ \frac 12 i g_{ab}\overline{\psi}^a  /\kern-.7em D
\psi^b+ \frac 18 (g_{ab} \overline{\psi}^a\psi^b)^2$,
where general principles and the previous coordinate choice yield
$D_\mu \psi^b= \partial_\mu \psi^b+ \Gamma^b_{cd}\partial_\mu \varphi^c
\psi^d$,  $~\Gamma^a_{bc}=
v^{aj}\partial^{\phantom{w}}_{(b}v_{c)}^{~~j}=\varphi^a g_{bc}$,
$~\partial^{\phantom{w}}_{[a} v_{~b]}^{~~j}=
\varepsilon^{jkl} v_a^{~k} v_b^{~l}$ ,
as well as
$g^{-1~ab}=\delta^{ab} - \varphi^a \varphi^b=v^{aj}v^{bj}$,
$v^{aj}= \sqrt{1-\varphi^2} ~\delta^{aj} + \varepsilon^{ajb} \varphi^b$.
Again, we rewrite ${\cal L}_{SCM}$ in terms of tangent space
quantities: the left-invariant tangent-space spinor is defined
(e.g.~\cite{geometrostasis}) by $\chi^j= v_a^j \psi^a$,
and transforms as $\delta \chi^j= \varepsilon^{jkl}\xi^k\chi^l$ under a full
$V+A$ transformation.  Thus, its contribution to the corresponding current
should be that of an isorotation. The tangent space lagrangean is
then\cite{geometrostasis}:
\begin{equation}
\label{LSCMtan}
{\cal L}_{SCM}=\frac 12 \!\left( J_\mu^j J^{\mu ~j}+
i \overline{\chi }^j /\kern-.5em\partial   ~\chi^j
+i\varepsilon^{jkl}\overline{\chi}^j /\kern-.5em J^k \chi ^l+
\frac 14 (\bar \chi^j \chi^j )^2 \right).
\end{equation}
This tangent space formulation is utilized below in bridging over to
the Supersymmetric Dual $\sigma$ Model (SDSM).

The supersymmetry transformations leaving the SCM action invariant are
$\delta \varphi^a = \bar \epsilon \psi^a $,
$~\delta \psi^a = ({\textstyle \frac 12} \Gamma^a_{bc} \overline{\psi}^b
 \psi^c  -i/\kern-.5em\partial  \varphi^a)\epsilon$.
In tangent space, these become
$ \delta J_\mu^j=-\bar\epsilon(\partial_\mu \chi^j + \varepsilon^{jkl} J_\mu
^k\chi^l)$, ~~$
\delta \chi^j= i/\kern-.5em J^j \epsilon -{\textstyle\frac 12}
\varepsilon^{jkl} (\gamma_p \epsilon ~\bar \chi^k \gamma_p \chi^l +
\gamma_\mu \epsilon ~ \bar \chi^k \gamma^\mu \chi^l ).$
Either form induces the conserved supercurrent\cite{witten}
\begin{equation}
S_\mu =-i/\kern-.5em \partial \varphi^a \gamma_\mu g_{ab} \psi^b~
= i /\kern-.5em J^j \gamma_\mu \chi^j ~.  \label{supercurrent}
\end{equation}
The curvature-free vector currents are again the bosonic pure gauges
$J$. However, the conserved vector currents now consist of
these bosonic terms augmented by spinor bilinears:
$C^i_\mu = J^i_\mu+K^i_\mu$, with
$J^i_\mu = -v_a^{~i} \partial_\mu \varphi^a, ~~
K^i_\mu =\frac i2\varepsilon^{ijk} \overline{\chi}^j \gamma_\mu \chi^k$.

Our new tangent space generator for a canonical transformation relating
$\varphi $ and $\chi$ at any fixed time to $\Phi $ and $X$
(the bosons and fermions of the dual theory) is
\begin{equation}
\label{F} F[\Phi,X ,\varphi,\chi] =
\int dx\;
\left(\Phi^j J^{1~j}[\varphi ]- \frac i2~\overline X^j \gamma^1 \chi^j \right)
\end{equation}
$=\int dx\Bigl(\Phi ^i (
\sqrt{1-{\varphi }^2}\frac {\stackrel{\leftrightarrow} {\partial}}
 {\partial x}\varphi ^i +\varepsilon
^{ijk}\varphi ^j\frac \partial {\partial x}\varphi ^k) -\frac i2
\overline X^j \gamma^1 \chi^j\Bigr)$.

Classically, the canonical conjugate to $\chi$
($\pi_{\chi} \equiv \delta {\cal L}_{SCM}/\delta \partial_0 \chi
= -i \chi^{\dag}/2$) is obtained from $F$ as
$-\delta F/\delta \chi = -iX^{\dag} \gamma_p /2 $, where
$\gamma_p=\gamma^0 \gamma^1$. So under the canonical transformation
\begin{equation}
\label{chiX}
\chi^j=\gamma_p X^j .
\end{equation}
Likewise, the momentum conjugate to $X$ is $\delta F/\delta X =
- i \chi^{\dag} \gamma_p/2$, leading to $\pi_X \equiv -i X^{\dag}/2$,
which specifies part of the dual lagrangean.
This chiral rotation of the fermions reflects the duality transition
of their bosonic superpartners, whose gradients map to curls
(in the weak field limit). The equal-time anticommutation
relations for Majorana spinors in tangent space,
$\{  \chi^j(x), \chi^k (y)\}=
\{  X^j(x), X^k (y)\}=2~ \delta^{jk} \delta (x-y)$, are preserved by the
above transformation, so it is properly identified as canonical.

Preservation of the canonical structure
for the bosons is less evident. The current algebra (i.e.~the Poisson
brackets for the $J$) is, indeed,  preserved in going over from the CM to the
DSM. However, preservation of the current algebra is necessary but {\em not}
sufficient for identification of such models\cite{balog}.
The classical conjugate momentum of $\Phi ^i$ is
\begin{eqnarray}
\label{Pi} \!\!
\Pi _i &=& \frac{\delta F}{\delta \Phi ^i}=
\left( \sqrt{1-{\varphi }^2}\;\delta ^{ij}+\frac{\varphi
^i\varphi ^j}{\sqrt{1-{\varphi }^2}}-\varepsilon ^{ijk}\varphi ^k\right)
\frac {\partial\varphi^j}{\partial x}=J^{1~i},
\end{eqnarray}
and the conjugate of $\varphi ^i$ is
\begin{eqnarray}
\label{PrelimVarPi}\!\!\!\!\!\!
\varpi _i&=&-\frac{\delta F}{\delta \varphi ^i}
=\left( \sqrt{1-
{\varphi }^2}\;\delta ^{ij}+\frac{\varphi ^i\varphi ^j}{\sqrt{1-{
\varphi }^2}}+\varepsilon ^{ijk}\varphi ^k\right) \frac {\partial\Phi^j}
{\partial x} +\left( \frac 2{\sqrt{1-{\varphi }^2}}\left( \varphi
^i\Phi ^j-\Phi ^i\varphi ^j\right) -2\varepsilon ^{ijk}\Phi ^k\right) \frac
\partial {\partial x}\varphi ^j.
\end{eqnarray}
These classical relations map the SCM as defined by
(\ref{LSCMtan}) to a dual theory, the SDSM.  The dual theory is
a supersymmetric extension of the bosonic
DSM, which contains {\em torsion}\cite{Fridling,CurtrightZachos'94},
and is defined by the following lagrangean.
\begin{eqnarray}
\label{LSDSM}
{\cal L}_{SDSM}&=&
\frac i2 \overline{X }^j /\kern-.5em\partial ~X^j
+\frac 18 (\overline X^j X^j )^2
-\frac 12 \Bigl( \varepsilon_{\mu\lambda}\partial^\lambda \Phi^a
-{\textstyle   \frac i2} \varepsilon^{ajk} \overline{X}^j \gamma_\mu X^k \Bigr)
 N^{\mu \nu}_{ab}
\Bigl( \varepsilon_{\nu\rho}\partial^\rho \Phi^b
-{\textstyle   \frac i2} \varepsilon^{blm}
\overline{X}^l \gamma_\nu X^m \Bigr)
\nonumber \\
&=&  \frac i2 \overline{X }^j /\kern-.5em\partial   ~X^j
+\frac i2 \varepsilon^{ijk} \overline {X}^i/\kern-.7em{\cal J}^j X^k
+\frac 38 (\overline X^i X^i ) \left( (\overline X^j X^j ) +
  { (\varepsilon^{jkl} \overline X^j \gamma_p X^k \Phi^l
+2 \overline X^j X^k  \Phi^j \Phi^k )\over 3~(1+4\Phi^2)} \right)
\nonumber \\
& &+\frac 12\left( G_{jk}~\partial _\mu \Phi^j\partial ^\mu \Phi^k
+ E_{jk}\varepsilon^{\mu\nu} \partial_\mu \Phi^j \partial_\nu \Phi^k \right),
\end{eqnarray}
where $N^{\mu \nu}_{ab} = g^{\mu \nu}G_{ab}+\varepsilon^{\mu \nu}~E_{ab}$,
$~G_{ab}=(\delta^{ab} + 4 \Phi^a \Phi^b)/(1+4\Phi^2)$,
$~E_{ab}=-2 \varepsilon_{abc} \Phi^c /  (1+4\Phi^2)$,
$G^{-1~ab}=(1+4\Phi^2)~ \delta^{ab} - 4 \Phi^a \Phi^b =V^{aj} V^{bj}$,
$~V^{aj}= \delta^{aj} - 2\varepsilon^{ajb} \Phi^b $, and
$~V^{~~j}_a=G_{aj} +E_{aj}$. Since
$\Phi_a=G_{ab} \Phi^b= \Phi^a$ and $\Phi^h= \Phi^a V_a^{~h}$,
it is unnecessary to distinguish the base- and
tangent-space indices
for the $\Phi$s.

The canonical momenta dictated by the lagrangeans ${\cal L}_{SCM}$ and
${\cal L}_{SDSM}$ are
\begin{equation}
\Pi _j=-    N^{1\nu}_{jk}\Bigl( \varepsilon_{\nu\rho}\partial^\rho \Phi^k
-{\textstyle   \frac i2} \varepsilon^{klm}
\overline{X}^l \gamma_\nu X^m \Bigr),
\end{equation}
\begin{eqnarray}
\varpi _i &=& \left( \delta ^{ij}+
\frac{\varphi ^i\varphi ^j}{1-{\varphi }^2}\right) \frac {\partial\varphi^j}
{\partial t} + (\sqrt{1-\varphi^2} \delta^{ij} +
{\varphi^i \varphi^j \over \sqrt{1-\varphi^2}}
+\varepsilon^{ijl}\varphi^l )~K_0^j [\chi]  .
\end{eqnarray}
We could now replace $\Pi$ and $\varpi$
with (\ref{Pi},\ref{PrelimVarPi})
to get covariant expressions for the dual transformations of fields.
Following that route leads to manifestly nonlocal expressions for
$\varphi$ and $\chi$ in terms of $\Phi$ and $X$.
Instead, we focus on the identification of the curvature-free currents in
the two theories, consistently with the above,
an identification which turns out to be completely local.

The crucial transition bridge to the SDSM relies on the bosonic current
\begin{eqnarray}
{\cal J}_j^\mu &=& -N^{\mu \nu}_{jk}
\varepsilon_{\nu\lambda}\partial^\lambda \Phi^k
=  \frac{-1}{1+4\Phi^2}\Bigl( (\delta ^{jl}+4\Phi^j\Phi^l)
\varepsilon ^{\mu \nu }\partial _\nu
\Phi ^l+2\varepsilon ^{jlk}\Phi ^l\partial ^\mu \Phi ^k\Bigr) ,
\end{eqnarray}
which conjoins with the fermionic bilinear component into
\begin{eqnarray}
\label{symphysis}
J^{\mu~j} &=&
-N^{\mu \nu}_{jk} ( \varepsilon_{\nu\lambda}\partial^\lambda \Phi^b
-{\textstyle   \frac i2} \varepsilon^{klm} \overline{X}^l \gamma_\nu X^m )
={\cal J}^{\mu j} + N^{\mu\nu}_{jk} K^k_{\nu}.
\end{eqnarray}
(Note from (\ref{chiX}), $K_\mu ^j [\chi] =K_\mu ^j [X]$.)
This pivotal, covariant, classical relation linking $J$ to $\Phi$ and
$X$ is derived in the canonical framework as follows.

The generating functional already automatically yielded the spacelike component
$J^1=\Pi={\cal J}^1 + (N\cdot K)^1$.
As in the bosonic model\cite{CurtrightZachos'94},
use of the variation (\ref{PrelimVarPi}) yields
a match for the timelike components as well,
by virtue of
\begin{equation}
\label{boscur}
-\sqrt{1-{\varphi }^2}\varpi_i-\varepsilon ^{ijk}\varphi ^j\varpi _k=
-\frac {\partial\Phi^i}{\partial x } - 2\varepsilon ^{ijk}\Phi ^j\Pi _k ,
\end{equation}
since both sides are equal to the mixed expression $-\partial\Phi^i/\partial x
+2\Phi ^j\left( \varphi^j \partial \varphi ^i /\partial x -\varphi ^i
\partial \varphi ^j /\partial x
\right)$   \linebreak
$+2\varepsilon ^{ijk}\Phi ^j\left( \varphi ^k \partial ( \sqrt{1-{\varphi }^2})
/\partial x -\sqrt{1-{\varphi }^2}\partial\varphi^k/\partial x \right)$.
Separation of bosonic from fermionic current pieces in Eq.~(\ref{boscur})
yields
$
J^i_0 -K_0^i={\cal J}_0^i - 2 \varepsilon^{ijk}\Phi^j (N\cdot K)^{k1}
$.
As a direct consequence,
\begin{equation}
J^0={\cal J}^0 + (N\cdot K)^0,
\end{equation}
so that the covariant identification of currents (\ref{symphysis}) holds
for the classical theory. Duality contrives to link  the SCM and SDSM
manifolds nontrivially. {\em Bosons in one theory are composites of
both bosons and fermions in the dual theory---a fermion-boson symphysis}.
Such mixings of boson and fermion bilinear components of currents are not
unfamiliar in standard canonical bosonization schemes\cite{Mandelstam}.

Predicated on the above identification of the vector currents,
the supercurrent (\ref{supercurrent}) for SDSM now simply reads
\begin{equation}
{\frak S}_\mu= i\gamma_p (/\kern-.7em{\cal J}^j + /\kern-.7em N\cdot
K^j  )\gamma_\mu X^j.
\end{equation}
Consequently, the respective supercharges identify in the two models, and
whence their squares, viz.~the hamiltonians. Likewise, by supertransforming
the supercurrents, the respective energy-momentum tensors identify,
including the respective hamiltonian densities. Hence, the SDSM is dynamically
equivalent to the SCM in phase space.
(Care must be taken in the identifications of covariant quantities, however,
such as lagrangean densities. Such identifications are direct only in terms of
canonical variables, and thus may only be valid ``on-shell'', since
supplanting $\pi$s with $\partial/\partial t$s requires Hamilton's equations.)

Naturally, the bosonic limits ($\chi=X=0$) of the above actions
are the models already connected in\cite{CurtrightZachos'94}. However,
the fermionic limits ($\varphi=\Phi=0$) are both the same Gross-Neveu model,
but with different normalizations of the interaction term,
$1/8\longrightarrow 3/8$,
respectively, as a consequence of the banished bosons in the above symphysis.
This change of the fermion coupling under such a singular transformation
evokes the change of coupling (i.e. inversion of radii)
under duality in bosonic models (on tori).

$F$ is ``covariant" with respect to supersymmetry transformations, in that
supertransforming its SCM variables is the same as supertransforming its SDSM
ones, essentially by construction of $\frak S$. Furthermore, the full
classical Poisson Bracket algebra of all generators ($F$, $H$, $Q$, etc.)
closes, but we will not discuss the details here.
We are more interested in the quantum algebra of the generators,
to which we now turn.

The generator $F$ actually relates the SCM to the SDSM not only at the
classical level, but also at the quantum level as well, just as in
the case of Liouville theory\cite{CurtrightGhandour}.
In general,  a classical generating
functional is the first step in Dirac's implementation\cite{Dirac33} of the
corresponding canonical transformation for the quantum theory.  The present
situation appears to be special, as was the purely bosonic
case\cite{CurtrightZachos'94}. We exponentiate the classical generating
functional to
obtain a candidate for the transformation functional connecting wave
functionals for the SCM to those for the SDSM:
\begin{equation}
T[\Phi,X ,\varphi,\chi] \equiv e^{ iF[\Phi,X ,\varphi,\chi]}.
\end{equation}
In a basis where the fields are diagonalized,
this $T$ is supposed to link states in the two theories by
\begin{equation}
\label{PAMD}
\langle \Phi, X \vert t \rangle = \int
T[\Phi,X ,\varphi,\chi]~\langle \varphi, \chi \vert t \rangle~d\varphi~d\chi~.
\end{equation}
The integrals on the RHS are over all field configurations at a fixed time,
$t$. It is in this sense that the two theories are canonically equivalent
at the quantum level\cite{Thornetal}.

We now argue that our candidate $T=\exp( {iF})$ provides exactly this link
by considering the action of local currents on the wave functionals.
It suffices to consider the vector currents and the supercurrents.
We show the action
of the SDSM currents on the SDSM wave functionals reduces to the action
of the corresponding SCM currents on the SCM wave functionals when
$T=\exp( iF)$ is used in (\ref{PAMD}).

This  follows through the use of integration by parts under the
functional integrals in (\ref{PAMD}), and the fact that
$T$ undergoes the same change when acted on by either set of currents.
We need only consider how the currents act on $T$, as explained for
the bosonic components of the currents in\cite{CurtrightZachos'94}.
There, as here, we express the classical currents in terms of fields and their
classical conjugate momenta, and then replace the classical conjugate momenta
with functional derivatives: $\varpi \to -i \delta/\delta \varphi$,
$\Pi \to -i \delta/\delta \Phi$. A similar procedure applies to the
spinors as well, with some care to maintain their Majorana
character.  Since $-i~\chi^{\dag}/2$ is the classical conjugate of $\chi$,
we would replace $\chi^{\dag} \to 2 \delta/\delta \chi$, given the
usual $\{\pi_{\chi}(x),\chi(y)\} = -i \delta (x-y)$.
But in the Majorana representation, $\chi^{\dag}=\chi^{T}$. Thus, a
prescription consistent with both is to replace
$\chi \to (\chi/\sqrt{2}~ + \sqrt{2}~ \delta/\delta \chi^{\dag})$
wherever the spinors appear in the currents. These prescriptions yield
currents as functional differential operators
acting on the space of wave functionals. It then follows that
\begin{eqnarray}
\label{superslick}
{\frak S}_{\mu} \langle \Phi, X \vert t \rangle
&=& \int~
\left({\frak S}_{\mu} e^{iF[\Phi,X ,\varphi,\chi]}\right)~\langle
\varphi, \chi \vert t \rangle~d\varphi~d\chi~
=\int~e^{iF[\Phi,X ,\varphi,\chi]}~
S_{\mu} \langle \varphi, \chi \vert t \rangle~d\varphi~d\chi~,
\end{eqnarray}
upon integrating by parts and discarding any surface terms in the
functional integration.
The steps in the calculation are the same as in the purely bosonic
case\cite{CurtrightZachos'94} with the additional identities
$(X^k + 2~ \delta/\delta X^{k~\dag}) \exp (\int dx\; \overline X^j
\gamma^1 \chi^j /2)=
(X^k + \gamma_p \chi^k)\exp (\int dx\; \overline X^j \gamma^1 \chi^j /2) =
\gamma_p~(\chi^k - 2~\delta/\delta \chi^{k~\dag})
\exp (-\int dx\;\overline \chi^j \gamma^1 X^j /2)$.
Finally, note $\delta/\delta \chi^{\dag}$ in the last term changes
sign upon integration by parts.  Results similar to (\ref{superslick})
hold for all the other local currents in the two models.
Thus, we conclude that our candidate transformation functional
induces the desired interchange of SDSM and SCM local currents acting
on the corresponding wave functionals.

Our argument for the quantum case is formal, obviously, since we have ignored
the UV divergences that inevitably arise in such manipulations.
To treat the problem with more care, we would need to discuss
the renormalization of $T$, $F$, and also the wave functionals
for both the SCM and the SDSM, in continuing parallel with the Liouville
case. This is beyond the scope of the present paper.
However, we may conjecture that the renormalization proceeds for $T$
in terms of the SCM manifold geometry by nothing but
the usual renormalization of the dreibein that appears in $F$, and
perhaps by rescaling (or at worst more complicated field redefinitions of)
the tangent space fields.  In general, we would also expect
the normalizations of the wave functionals to be changed
by the transformation but only through energy-dependent factors.
If this is indeed so, the formal arguments
above have given the correct structure of the exact answer, just as for
the Liouville theory.

\acknowledgements
Work supported by the NSF grant PHY-92-09978 and
the U.S.~Department of Energy, Division of High Energy Physics, Contract
W-31-109-ENG-38. C.~Zachos is obliged to the Yukawa Institute for Theoretical
Physics for its gracious hospitality during the inceptive phase of this study,
and especially to T.~Uematsu, without whose incisive questions these results
would not have been sought.   T.~Curtright wishes to thank O.~Alvarez
for communicating his results on duality and tangent space prior to
publication.

\end{document}